\documentclass[a4paper]{article}
\usepackage{graphicx}
\def\putps(#1,#2)(#3,#4)#5#6{\ifnum\Mac=1 \put(#1,#2){\special{picture #5}}
\else  \put(#3,#4){\includegraphics{#6}} \fi}

\newcommand{\U}{\,{\rm U(1)}}
\newcommand{\GeV}{\,{\rm GeV}}

\newcommand{\fig}[1]{~\ref{fig:#1}}
\newcommand{\lascia}[1]{}

\font\tenrsfs=rsfs10
\font\sevenrsfs=rsfs7
\font\fiversfs=rsfs5
\newfam\rsfsfam
\textfont\rsfsfam=\tenrsfs
\scriptfont\rsfsfam=\sevenrsfs
\scriptscriptfont\rsfsfam=\fiversfs
\def\mathscr#1{{\fam\rsfsfam\relax#1}}
\def\Lag{\mathscr{L}}

\usepackage{graphicx}
\usepackage{multicol}
\usepackage{color}

\definecolor{rosso}{cmyk}{0,1,1,0.4}
\definecolor{rossos}{cmyk}{0,1,1,0.55}
\definecolor{rossoc}{cmyk}{0,1,1,0.2}
\definecolor{blu}{cmyk}{1,1,0,0.3}
\definecolor{blus}{cmyk}{1,1,0,0.6}
\definecolor{bluc}{cmyk}{1,1,0,0.1}
\definecolor{verde}{cmyk}{0.92,0,0.59,0.25}
\definecolor{verdec}{cmyk}{0.92,0,0.59,0.15}
\definecolor{verdes}{cmyk}{0.92,0,0.59,0.4}
\definecolor{giallo}{cmyk}{0,0,1,0}
\definecolor{gialloverde}{cmyk}{0.44,0,0.74,0}
\definecolor{purple}{cmyk}{0.44,1,0.74,0}

\newcommand{\TeV}{\,{\rm TeV}}

\newcommand{\NP}{Nucl. Phys.}

\newcommand{\PRL}{Phys. Rev. Lett.}
\newcommand{\PL}{Phys. Lett.}
\newcommand{\PR}{Phys. Rev.}

\newcommand{\eq}[1]{~(\ref{eq:#1})}

\newcommand{\hc}{\hbox{h.c.}}

\def\Emisst{E_T\hspace{-2.6ex}/\hspace{2ex}}
\def\Emiss{E\hspace{-1.3ex}/\hspace{0.3ex}}
\def\beq{\begin{equation}}
\def\eeq{\end{equation}}
\def\bea{\begin{eqnarray}}
\def\eea{\end{eqnarray}}

  \def\SU{{\rm SU}}
 
\def\circa#1{\,\raise.3ex\hbox{$#1$\kern-.75em\lower1ex\hbox{$\sim$}}\,}
\makeatletter
\def\art{\@ifnextchar[{\eart}{\oart}}
\def\eart[#1]#2#3#4#5#6{{\rm #2}, {\em #3 \rm #4} {\rm (#6) #5 ({\em #1})}}
\def\hepart[#1]#2{{\rm #2, \em#1}}
\newcommand{\oart}[5]{{\rm #1}, {\em #2 \rm #3} {\rm (#5) #4}}

\newcounter{alphaequation}[equation]
\def\thealphaequation{\theequation\hbox to
0.6em{\hfil\alph{alphaequation}\hfil}}
\def\eqnsystem#1{
\def\@eqnnum{{\rm (\thealphaequation)}}
\def\@@eqncr{\let\@tempa\relax \ifcase\@eqcnt \def\@tempa{& & &} \or
  \def\@tempa{& &}\or \def\@tempa{&}\fi\@tempa
  \if@eqnsw\@eqnnum\refstepcounter{alphaequation}\fi
\global\@eqnswtrue\global\@eqcnt=0\cr}
\refstepcounter{equation} \let\@currentlabel\theequation \def\@tempb{#1}
\ifx\@tempb\empty\else\label{#1}\fi
\refstepcounter{alphaequation}
\let\@currentlabel\thealphaequation
\global\@eqnswtrue\global\@eqcnt=0 \tabskip\@centering\let\\=\@eqncr
$$\halign to \displaywidth\bgroup \@eqnsel\hskip\@centering
$\displaystyle\tabskip\z@{##}$&\global\@eqcnt\@ne
\hskip2\arraycolsep\hfil${##}$\hfil& \global\@eqcnt\tw@\hskip2\arraycolsep
$\displaystyle\tabskip\z@{##}$\hfil
\tabskip\@centering&\llap{##}\tabskip\z@\cr}
\def\endeqnsystem{\@@eqncr\egroup$$\global\@ignoretrue} \makeatother

\begin{document}
\centerline{hep-ph/0301232\hfill
IFUP--TH/2003-4 \hfill   CERN-TH/2003-013}

\vspace{1cm}
\centerline{\LARGE\bf Constraints on Extra-Dimensional Theories}
\vspace{0.3cm}
\centerline{\LARGE\bf from Virtual-Graviton Exchange}\vspace{3mm}
\bigskip\bigskip

\centerline{\large\bf Gian Francesco Giudice}\vspace{0.1cm}
\centerline{\em Theoretical Physics Division,  CERN, CH-1211, Gen\`eve 23, Switzerland}\vspace{0.3cm}
\centerline{\large\bf Alessandro Strumia}\vspace{0.1cm}
\centerline{\em Dipartimento di fisica dell'Universit\`a di Pisa and INFN, 
Italy}

\bigskip\bigskip
\centerline{\large\bf Abstract}\begin{quote}\large\indent
We study the effective interactions induced by loops of extra-dimensional
gravitons and show the special r\^ole of a specific dimension-6 four-fermion
operator, product of two flavour-universal axial currents. By introducing
an ultraviolet cut-off, we compare the present constraints on low-scale
quantum gravity from various processes involving real-graviton emission
and virtual-graviton exchange. The LEP2 limits on dimension-6 four-fermion
interactions give one of the strongest constraint on the theory, in particular
excluding the case of {\it strongly-interacting} gravity at the weak scale.
\end{quote}
\noindent

\section{Introduction}

In this paper we consider scenarios in which the Standard Model fields are
confined on a $3$-dimensional brane, while gravity propagates in the full 
$D$-dimensional space, with
$\delta$ flat and compactified
extra spatial dimensions ($D=4+\delta$)~\cite{add}. 
Since the theory is non-renormalizable, it lacks much
predictivity in the quantum domain, as a consequence
of our ignorance of the short-distance regime. 
Nevertheless, at energies much smaller 
than $M_D$ (the 
Planck mass of the $D$-dimensional theory) we can use an effective
theory to correctly describe the interactions of the extra-dimensional
gravitons with gauge and matter fields~\cite{noi,pesk,han}. 
The effective theory, 
in particular,
allows the calculation of the rate for soft-graviton emission in high-energy
colliders. Also, at energies much larger than $M_D$, one can use a
semi-classical approach to compute gravitational scattering with small
momentum transfer~\cite{trans} and estimate black-hole production~\cite{bh}.

Observables dominated by ultraviolet contributions 
are experimentally interesting but
remain unpredictable,
lacking a knowledge of the underlying theory at short distances. 
In particular this is the case for effects induced by virtual graviton exchange,
both at tree and loop levels. A general, but not very restrictive,
parametrization of all virtual effects 
can be done in terms of higher-dimensional effective
operators. Many theoretical and experimental analyses 
have concentrated on the dimension-8
operator $T_{\mu\nu}T^{\mu\nu}$
induced by tree-level graviton exchange~\cite{noi,han,hewett}.

Here we want to describe a parametrization which, to a certain extent,
allows a comparison among different observables induced by graviton
quanta~\cite{Contino}. 
This requires the
introduction of a new mass scale, besides the
fundamental parameters $M_D$ and $\delta$, which we will call $\Lambda$
and which represents the validity cut-off scale of the Einstein gravitational
action. As an analogy, consider 
the case of the Standard Model, in which the complete theory is known and 
perturbative. If we compare radiative corrections computed in the full 
theory with the results obtained by cutting off power-divergent loops
in the Fermi theory, we find that the appropriate cut off is given by
the smaller scale
$\Lambda \sim M_W \approx g G_{\rm F}^{-1/2}$
rather than by the larger Fermi scale $\Lambda \approx G_{\rm F}^{-1/2}$.
Similarly, we will estimate graviton-loop effects by introducing an 
explicit ultraviolet cutoff $\Lambda$, that can be different
from $G_D^{1/(2-D)}$, where $G_D$ is the $D$-dimensional Newton constant.
 We expect that $\Lambda$
represents the mass of the new states introduced by the short-distance
theory to cure the ultraviolet behavior of gravity. 
However, for our purposes,
$\Lambda$ is just a phenomenological parameter.
From the point of view of the effective theory,
 $\Lambda/M_D$ parametrizes how strongly (or weakly) coupled quantum
gravity is, and therefore controls the unknown relative importance of 
tree-level versus loop graviton effects.
  
We should immediately emphasize that the cut-off procedure is arbitrary
and therefore the comparison between different observable has, at
best, a semi-quantitative meaning. We will discuss how the ``geometrical''
factors in front of the divergent loop integrals can be estimated with
the language of na\"{\i}ve dimensional analysis~\cite{NDA} and we will
compare them with the results obtained with specific
regulators. Of course, order-one coefficients 
are unpredictable, and therefore our results are only to be viewed
as ``reasonable estimates''.

Nevertheless, some interesting results can be obtained. First of all,
even in a weakly-coupled theory, graviton loops can be more important
than tree-level graviton exchange. This is because tree-level effects
generate a dimension-8 operator, while loop effects can generate operators
of lower dimension. Secondly, using the symmetry properties of gravity,
we will show that graviton loops can generate only a single 
dimension-6 operator involving fermions, here called $\Upsilon$,
which is the product of two
flavour-universal axial-vector currents, see eq.~(\ref{upsilon}). 
This operator, in conjunction with 
graviton emission and the $T_{\mu\nu}T^{\mu\nu}$ operator, should be the
object of analysis in complete experimental studies of graviton effects
at colliders. Indeed, we will show that the LEP2 limit on $\Upsilon$
at present provides one of the  strongest constraint on extra-dimensional
gravity, in particular excluding the case of a fully strongly-interacting
theory within the range of near-future colliders.
Finally, graviton loops can also generate
some other dimension-6 operators relevant only for Higgs physics. 

This paper is organized as follows. In sect.~2 we discuss different approaches
to estimate divergent loops. Then, in sects.~3 to 6 we compute the effects
from graviton emission, graviton exchange at tree and loop level, and
loops with both graviton and gauge bosons. The different constraints
are compared in sect.~7 and our conclusions are presented in sect.~8.

\section{Dealing with divergences}\label{2}

We start by discussing the procedure we follow to estimate divergent
loop coefficients.
Na\"{\i}ve Dimensional Analysis (NDA)~\cite{NDA} represents a successful way 
to estimate the geometrical factors multiplying the coefficients
of effective operators generated by strongly-interacting dynamics.
The rules of NDA dictate that, working in units of the strongly-interacting
scale $\Lambda_{\rm S}$, the value of the dimensionless coupling $g$ of the
interaction (at the scale $\Lambda_{\rm S}$) is such that any loop order
gives an equal contribution. Each additional loop carries a factor
$g^2/(\ell_4 \ell_\delta)$, where the $D$-dimensional loop factor is
defined as\footnote{Starting from the equation
$$
\int \frac{d^Dk}{(2\pi)^D} ~f(k^2)=\frac{S_{D-1}}{2(2\pi)^D}
\int dk^2~(k^2)^{\frac{D}{2}-1} f(k^2),
\nonumber
$$
where $S_{D-1}=2\pi^{D/2}/\Gamma (D/2)$ 
is the surface of the unit-radius sphere in $D$ dimensions,
we define as $1/\ell_D$ the coefficient of the integral in the right-hand
side. We find $\ell_D= 2^D \pi^{D/2}(\frac{D}{2}-1) ! $ for $D$ even,
and $\ell_D= 2^D \pi^{(D+1)/2} [\Pi_{k=0}^{(D-3)/2} (k+1/2)]$ for $D$ odd.}
\beq
\ell_D=(4\pi)^{D/2} \Gamma(D/2).
\eeq
Notice that, for simplicity, we have chosen to factorize the $D$-dimensional
loop integration in the usual 4-dimensional coordinates (with 
$\ell_4=16\pi^2$) and in the $\delta$ extra coordinates (where the
integration actually corresponds to the summation over the Kaluza-Klein
modes of the compactified space).
Then the value of $g^2$ that corresponds to a strongly-coupled interaction is
given by 
\beq
g^2 = g^2_\star \equiv \ell_4 \ell_\delta.
\eeq

The coupling of the $D$-dimensional graviton in the effective 
theory can be written as 
\beq
g^2=c_\delta (E/M_D)^{2+\delta}, ~~~c_\delta \equiv (2\pi )^\delta , 
\eeq
where
$E$ is the typical energy and
$c_\delta$ is the coefficient relating the $D$-dimensional
Planck mass to the reduced Planck mass, given here following
the notations of ref.~\cite{noi}. 
Therefore the energy scale where gravity becomes strong, as defined by NDA, is
\beq
\Lambda_{\rm S} = \left( \frac{\ell_4 \ell_\delta}{c_\delta}\right)^{\frac{
1}{2+\delta}} M_D 
= \left[16 \pi^{2-\frac{\delta}{2}}~
\Gamma\left(\frac{\delta}{2}\right) 
\right]^{\frac{
1}{2+\delta}} M_D .
\label{strongl}
\eeq

As mentioned before, we are interested in the possibility that
the ultraviolet behavior of gravity is modified before the theory becomes 
strongly interacting. In this case, the NDA estimates can still be used 
by taking $g^2=\epsilon g_{\star}^2$, where $\epsilon$ is a small parameter
that measures the weakness of the couplings~\cite{luty}. If we 
parametrize this weakness by introducing an ultraviolet cut off $\Lambda$,
we can take $\epsilon =(\Lambda /\Lambda_{\rm S} )^{2+\delta}$. 
Therefore, the coefficient of any desired operator is estimated by multiplying
the appropriate power of the graviton coupling $(c_\delta/M_D^{2+\delta})^{1/2}$
times the appropriate 4-dimensional and extra dimensional loop factors
$\ell_4$ and $\ell_\delta$,
times the appropriate power of $\Lambda$ (and not $\Lambda_{\rm S}$)
needed to match the
correct dimensions of the effective operator.
Concrete examples are discussed in the next sections,
where we will use this ``modified
NDA''  to estimate the effects of virtual graviton exchange and to make a
``reasonable'' comparison between virtual and real graviton effects. 

In the following we will also compare the results from ``modified NDA''
with the results obtained from explicit perturbative calculations
with ultraviolet regulators. One possibility we consider is to cut-off
each divergent integral at a common scale $\Lambda$. This procedure
should somehow reproduce the appearance of the new states with mass
$\Lambda$ that tame the ultraviolet behavior of the theory. Another
example we study is to multiply each divergent integrand by a factor
$\exp (-\Lambda^2/k^2)$, and integrate over the full range of internal
momenta $k$. This could mimic the exponential suppression at a string
scale of order $\Lambda$. Of course these procedures, which are not
unambiguously defined and in particular break gauge invariance, can
only be viewed as tools to obtain estimates and order-unity factors
in the results cannot be trusted. Moreover, the cutoff parameter is not 
necessarily universal,
i.e.\ contributions to different processes might be cut off 
by different values of $\Lambda$.
 
Another procedure we will follow is to assume that the graviton exchange
(at tree- or loop-level) mediates interactions between two currents
located at different points along the extra-dimensional coordinates.
In such a case, the coefficient of any contact operator involving different
particles separated by a distance $r$ 
is finite and computable.
This is because quantum gravity,
despite being non-renormalizable,
is a local theory, and non-local operators cannot receive 
ultraviolet-divergent corrections.
The divergence is recovered by letting $r\to 0$;
therefore we can use the splitting as a physical regulator
by identifying $r^{-1}$ with the cut-off $\Lambda$.
This procedure is well-defined and gauge-invariant, and it corresponds to the
assumption that different matter fields are located on different branes
in the extra-dimensional space. This hypothesis was proposed in
ref.~\cite{schmaltz} 
in order to alleviate the problems of baryon-number
and flavour-symmetry violations and to understand the fermion-mass
pattern. However, this procedure does not regularize graviton contributions
to multiple self-interactions of the same matter field, {\it e.g.}
like effective operators mediating Bhabha scattering.

\begin{figure}[t]
\centering
\includegraphics[width=0.9\linewidth]{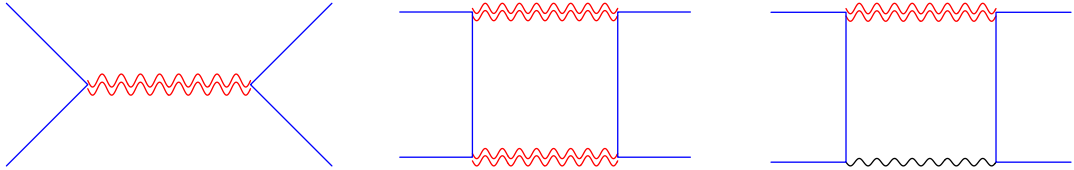}
\caption{\em Fig.~\ref{fig:treeloop}{\rm a}:
tree-level graviton exchange generating the dimension-8 operator $\tau$.
Fig.~\ref{fig:treeloop}{\rm b}: one-loop graviton exchange generating 
the dimension-6 operator $\Upsilon$.
Fig.~\ref{fig:treeloop}{\rm c}: exchange of gravitons and vector bosons 
at one loop generating dimension-6 operators that affect electroweak precision
data.
\label{fig:treeloop}}
\end{figure}

\section{Graviton emission}
  
Graviton emission in particle scattering can be computed using the
infrared properties of gravity. Therefore its theoretical prediction
is independent of $\Lambda$, as long as the relevant energy of the 
process is smaller than $\Lambda$. Checking the validity of this
assumption, necessary for employing the effective-theory result, 
is non-trivial in the case of hadron colliders, where parton scattering
can occur at very different center-of-mass energies. 
Applying the
method suggested in ref.~\cite{noi}, one can define a 
minimum value of $M_D$, for a given $\delta$, below which the
effective theory cannot be trusted. For instance, 
the window in
$M_D$ values where graviton emission is observable at the LHC and
is reliably estimated by the theory
disappears for $\delta \ge 5$.
Graviton emission rates in LHC collisions are suppressed
if $\Lambda$ is smaller than the LHC center of mass energy.

Graviton emission has been studied in the processes $e^+e^-\to
\gamma \Emiss$ and  $e^+e^-\to
Z \Emiss$ at LEP, and $p\bar p \to {\rm jet} + \Emisst$ and
$p\bar p \to \gamma + \Emisst$ at the Tevatron. The present limits
at 95\% CL are summarized in table~\ref{tab1}\footnote{We
do not include a CDF analysis~\cite{cdf} on single-photon events, since the 
obtained limits on $M_D$ grow with $\delta$, signalling that the relevant
kinematics are probably outside the validity range of the effective theory.},
together with the combined bound.
We have combined the bounds from the different experiments assuming that, 
whenever the experiment does not report the
$\chi^2$ dependence, the best-fit value occurs
at $M_D \to \infty$. Therefore, our procedure is only approximate.
At any rate, the effect of the combination gives
only a moderate improvement of the strongest single bound.

\begin{table}
\centering
\begin{tabular}{|c|ccccc|}
\hline\hline
Experiment & $\delta =2$ & $\delta =3$ & $\delta =4$ & $\delta =5$ 
& $\delta =6$ \\
\hline
ALEPH \cite{aleph} &1.26&0.95 &0.77&0.65&0.57\\
DELPHI \cite{delphi} &1.36&1.05 &0.84&0.69&0.59\\
L3 \cite{l3} &1.02 &0.81&0.67&0.58&0.51\\
OPAL \cite{opal} &1.09 &0.86&0.71&0.61&0.53\\
D{\O} \cite{d0} &0.89 &0.73&0.68&0.64&0.63\\
\hline
combined &1.45&1.09&0.87&0.72&0.65\\
\hline\hline
\end{tabular}
\caption{\label{tab1}
\em {95\% CL limits on the $D$-dimensional Planck mass 
$M_D$ (in TeV), for some values
of the number of extra dimensions $\delta$, from {\em graviton-emission}
processes in different
experiments.
}}
\end{table}

\section{Virtual graviton exchange at tree level}
Tree-level exchange of gravitons generates the effective dimension-8 
operator $\tau$~\cite{noi,han,hewett} (see fig.~\ref{fig:treeloop}a)
\beq
\Lag_{\rm int} =c_\tau ~\tau, ~~~~~~
\tau=\frac{1}{2}\left( T_{\mu\nu}T^{\mu\nu} -\frac{T_\mu^\mu T_\nu^\nu}{\delta +2} 
 \right) ,
\label{tau}
\eeq
where $T_{\mu \nu}$ is the energy-momentum tensor.
Its coefficient $c_\tau$
is divergent because, although we are dealing with a tree-level
contribution, we have to sum over all possible intermediate graviton
configurations that conserve 3-dimensional momentum and energy, but not
necessarily the extra-dimensional momentum. Therefore $c_\tau$
depends on the cutoff $\Lambda$, and it cannot be computed from 
the low-energy effective theory.
We 
can estimate it by using the rules of ``modified NDA'' presented before:
\beq
c_\tau^{\rm NDA} = \frac{c_\delta \Lambda_{\rm NDA}^{\delta -2}}{\ell_\delta 
M_D^{\delta +2}}
=\frac{\pi^{\delta /2}}{\Gamma(\delta /2)}\frac{\Lambda_{\rm NDA}^{\delta -2}}
{M_D^{\delta +2}}.
\label{estau}
\eeq
We have introduced here a subscript in the definition of the
cut-off $\Lambda_{\rm NDA}$ to distinguish it from the cut-off parameter
used in the different regularization procedures that we will now study.

An explicit calculation of $c_\tau$ gives 
\beq 
c_\tau = \frac{c_\delta}{M_D^{\delta +2}} G(k\to 0,r=0)
\eeq
where $G(k,r)$ is the scalar propagator 
with momentum $k$ between two points separated by a distance 
$r\equiv |\vec r|$ in the extra dimensions:
\beq
G(k,r) = \frac{1}{V}\sum_{\vec n}
\frac{e^{i  \vec{n}\cdot\vec{r}}/R}{k^2+|\vec n|^2/R^2}.
\eeq
Here $V$ is the volume of the extra dimensions.
For simplicity we assume a toroidal compactification, so that $V=(2\pi R)^\delta$ where $R$ is the radius.
For sufficiently large extra dimensions, we can convert the sum into
an integral over a continuous Kaluza-Klein mass distribution (with $m=
|\vec n|/R$) and over the angle $\theta$ formed by the two vectors $\vec n$
and $\vec r$:
\beq
G(k,r)= \frac{S_{\delta -2}}{c_\delta} \int_0^\infty dm~
m^{\delta -1} \int_0^\pi d\theta ~\sin^{\delta -2} \theta
~~\frac{e^{imr\cos\theta}}{k^2+m^2}.
\label{gfun}
\eeq

For $r\ne 0$, we can perform the integrals in eq.~(\ref{gfun}) and find
\beq
G(k,r)=\frac{1}{(2\pi)^{\delta /2}} \left( \frac{k}{r}\right)^{\frac{\delta}{2}
-1} K_{\frac{\delta}{2}-1}(kr) ~~~\stackrel{kr \to 0} {\simeq}~~~
\frac{\Gamma \left( \frac{\delta}{2}-1\right) r^{2-\delta}}{4\pi^{\delta /2}},
\label{gfunn}
\eeq
where $K_\nu$ is a modified Bessel function.

If we take $r=0$, we obtain
\beq
G(k,0)=\frac{S_{\delta -1}}{2c_\delta} \int_0^\infty dm^2
\frac{(m^2)^{\frac{\delta}{2}-1}}{k^2+m^2}.
\label{gfunz}
\eeq
For $\delta \ge 2$, the integral is divergent and must be regularized by
some {\it ad hoc} procedure.
We can introduce an unknown function $f(m^2/\Lambda^2)$
that acts as an ultraviolet regulator at the scale $\Lambda$.
From eq.~(\ref{gfunz}) we find
\beq
c_\tau =c_\tau^{\rm NDA} \left( \frac{\Lambda}{\Lambda_{\rm NDA}}\right)^{\delta
-2}
\int_0^\infty dx~f(x)~x^{\frac{\delta}{2} -2},
~~~x\equiv m^2/\Lambda^2.
\eeq
\begin{itemize}
\item[a)]
If we choose $f(x)=e^{-x}$, as proposed in sect.~2,
we obtain
\beq
\frac{c_\tau}{c_\tau^{\rm NDA}}
\left( \frac{\Lambda_{\rm NDA}}{\Lambda} \right)^{\delta -2}
= \frac{2\Gamma(\delta/2)}{\delta -2}.
\label{tred}
\eeq
The right-hand side of eq.~(\ref{tred}) 
varies between $\sqrt{\pi}$ and 1, for $\delta$
between 3 and 6. 

\item[b)] If we assume a sharp cut-off at the scale $\Lambda$ and take $f(x)=1$
for $x<1$ and $f(x)=0$ otherwise, we obtain
\beq
\frac{c_\tau}{ c_\tau^{\rm NDA}} 
\left( \frac{\Lambda_{\rm NDA}}{\Lambda} \right)^{\delta -2}
=  \frac{2}{\delta -2}.
\label{quatt}
\eeq
Equation~(\ref{quatt}) varies between 2 and 1/2, for $\delta$
between 3 and 6. Therefore, for both choices of regulators a) and b), 
the results
for $c_\tau$ (with $\delta \le 6$) are in fair agreement with the 
estimates using ``modified NDA'', for equal values of $\Lambda$ and
$\Lambda_{\rm NDA}$.

\item[c)]
Finally, we assume that different matter particles interacting
with the gravitons are localized
at different points in the extra dimensions.
Using eq.~(\ref{gfunn}) and
identifying $r=1/\Lambda \ll R$, we find
\beq 
\frac{c_\tau }{ c_\tau^{\rm NDA} }
\left( \frac{\Lambda_{\rm NDA}}{\Lambda} \right)^{\delta -2}
 = \frac{2^{\delta -1}\Gamma^2(\delta/2)}{\delta -2}.
\label{raton}
\eeq
Here, with an abuse of notation, we have
denoted by $c_\tau$ the terms in the operator $\tau$ involving splitted
particles. The right-hand side of 
eq.~(\ref{raton}) grows fast with $\delta$ and
it is already equal to 32 for $\delta=6$.

\end{itemize}
In the case $\delta =2$, there is an infrared logarithmic divergence
and, for
all kinds of regulators considered here, 
$c_\tau$ becomes
\beq
c_\tau = \frac{\pi}{M_D^4} \ln \frac{\Lambda^2}{E^2},
\eeq
where $E$ is the typical energy exchanged in the process.

The operator in eq.~(\ref{tau}) does not affect precision observables 
at the  $Z$-resonance, but it gives anomalous contributions to
many ordinary particle processes. At LEP, the most sensitive channels
are Bhabha scattering and diphoton productions, but experiments have
also set limits on $c_\tau$ from dilepton, dijet, $b\bar b$, $WW$,
and $ZZ$ production. Experiments at the Tevatron have set constraints on
$c_\tau$ from studying dielectron and diphoton final states. Experiments
at HERA have also provided constraints from $e^\pm p \to e^\pm p$.
Present bounds on $c_\tau$ are summarized in 
table~\ref{tab2}\footnote{In the notations of refs.~\cite{noi} and
\cite{hewett}, we find $c_\tau= 4\pi/\Lambda_T^4 =8/M_S^4$. 
We do not use the strong limit from the 
$ZZ$ channel preliminarily reported by L3, since this result has not been
confirmed by the full analysis.}. We have combined different experiments
following the same approximation used in the previous section.

\begin{table}
\centering
\begin{tabular}{|c|c|cc|}
\hline\hline
Experiment&Process & + & $-$\\ 
\hline
LEP combined~\cite{lepc} & $\gamma \gamma$ & 0.93&1.01\\
LEP combined~\cite{lepcc} & $e^+e^-$ & 1.18 & 1.17\\
H1~\cite{h1} & $e^+p$ and $e^-p$ & 0.74 & 0.71\\
ZEUS~\cite{zeus} & $e^+p$ and $e^-p$ & 0.72 & 0.73\\
D\O~\cite{d0c} & $e^+e^-$ and $\gamma \gamma$ & 1.08 & 1.01\\
CDF~\cite{zhou} &  $e^+e^-$ and $\gamma \gamma$ & 0.86 & 0.84\\
\hline
\multicolumn{2}{|c|}{combined}&1.28&1.25\\
\hline\hline
\end{tabular}
\caption{\label{tab2}
{\em Dimension-8 operator $\tau$}:
\em {95\% CL limits on $|8/c_\tau|^{1/4}$ (in {\rm TeV}) for positive
and negative values of $c_\tau$,
from different
experiments. 
}}
\end{table}

\begin{figure}[p]
\vspace{-1cm}
\centering
\includegraphics[width=0.8\linewidth]{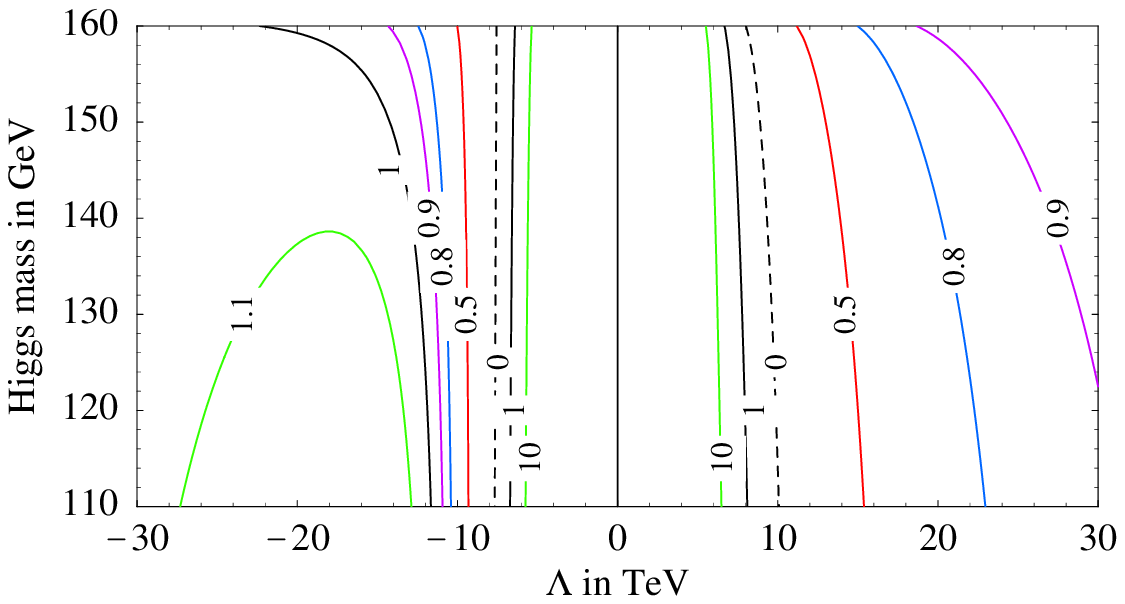}
\caption{\em 
$\sigma (gg\to h)\times {\rm  
BR}(h\to \gamma \gamma)$ in units of its SM value, in presence of the
new operators $-(\pi/\Lambda^2)H^\dagger H F_{\mu\nu}F^{\mu\nu}$ and
$-(\pi/\Lambda^2)H^\dagger H G^a_{\mu\nu}G^{a\mu\nu}$.
 \label{fig:higgs}}
\vfill
\includegraphics[width=0.85\linewidth]{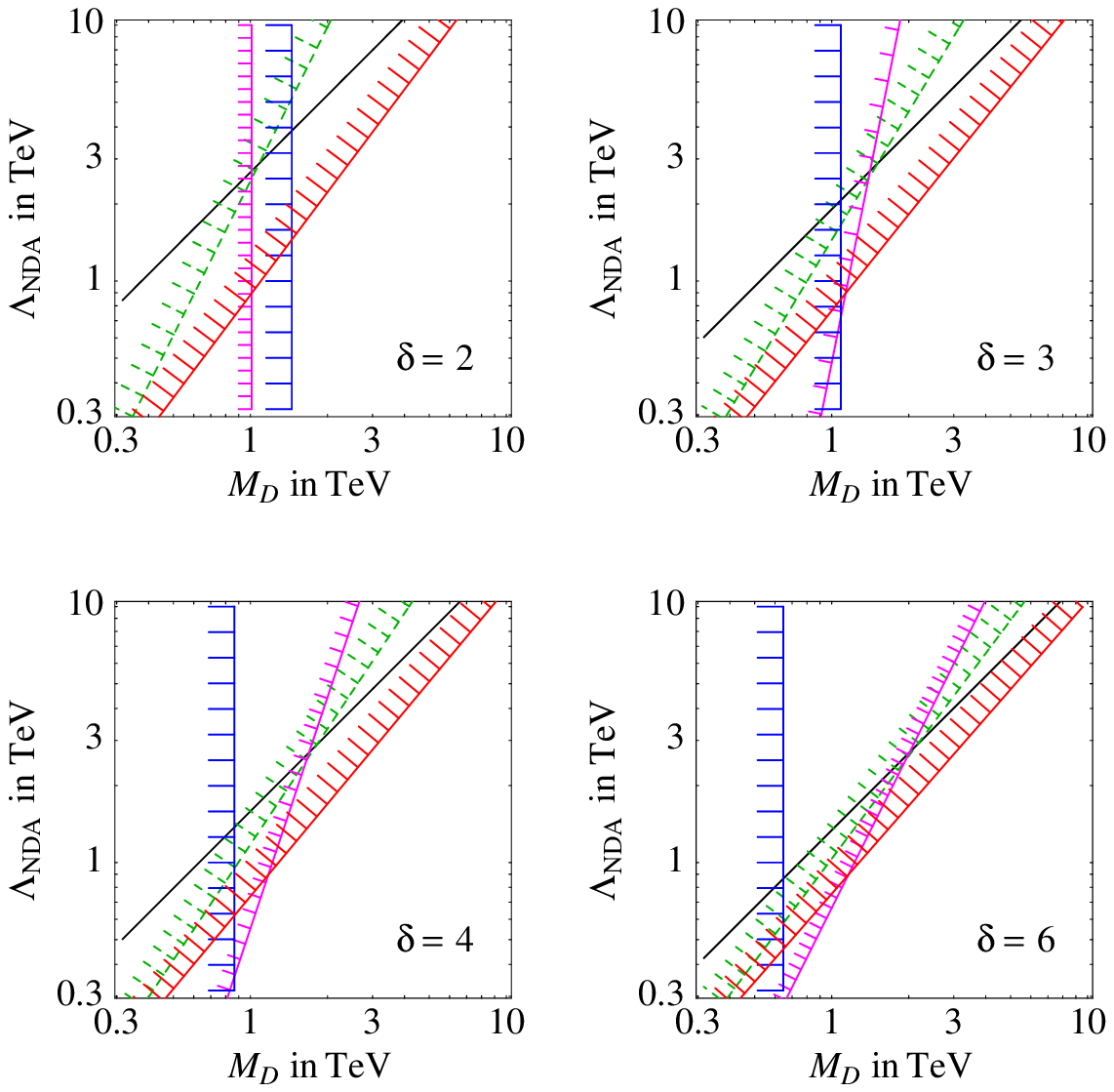} 
\caption{\em  
95\% CL collider bounds on graviton phenomenology in 
the plane $(M_D,\Lambda_{\rm NDA})$ for
$\delta=2,~3,~4,~6$ flat extra dimensions.
The solid black line 
shows the value
of the cut-off $\Lambda_{\rm NDA}$ corresponding to a strongly-interacting
gravitational theory, as defined from NDA, see eq.~(\ref{strongl}).
The other lines mark the regions excluded by the bounds from 
{\color{blus} graviton emission (vertical blue line)}, 
{\color{purple} tree-level virtual graviton exchange (purple line 
with short borderlines)}, 
{\color{rossos} graviton loops (red solid line)}, 
{\color{verdes} graviton and gauge boson loops (green dashed 
line)}.\label{fig:vincoli}}
\end{figure}

\section{Graviton loops}
Loops with exchange of virtual gravitons,
like those illustrated in fig.~\ref{fig:treeloop}b, can be important
because  generate operators with 
dimensionality lower than 
the operator in eq.~(\ref{tau}), arising from tree-level graviton
exchange. If we consider only fermions or gauge bosons, 
a single operator with dimension less than 8 can be generated, and it
is given by
\beq
\Lag_{\rm int} =c_\Upsilon ~\Upsilon, ~~~~~~
\Upsilon =\frac{1}{2} \left( 
\sum_{f=q,\ell}\bar f \gamma_\mu \gamma_5 f \right)
\left( 
\sum_{f=q,\ell}\bar f \gamma^\mu \gamma_5 f \right) .
\label{upsilon}
\eeq
Here the sum extends over all quarks and leptons in the theory. It is
easy to realize that graviton loops cannot produce any
other operator of dimensions 6 or 7, involving
only fermions and gauge bosons.
Since the intermediate state involves virtual gravitons, any
induced operator should be the product of two currents, which are singlets
under all gauge and global symmetry groups, and are even under charge
conjugation. 

The operator $\Upsilon$ in eq.~(\ref{upsilon}) is already
present in the  tree-level  gravitational Lagrangian, if one employs a formalism
in which the connection is not symmetric so that there is a torsion coupled to the spin density of
matter.
Within a given formalism its coefficient is not computable 
in the context of extra dimensions with matter localized on a 
infinitesimally thin brane~\cite{takeuchi}.
In the case of supergravity
with minimal K\"ahler potential, $\Upsilon$ is not present in the tree-level
Lagrangian. In the context of extra dimensions, $\Upsilon$ 
was previously discussed in ref.~\cite{takeuchi}.
Here we make the conservative assumption that the coefficient
of the operator $\Upsilon$ vanishes at tree level and we estimate its
contribution from quantum loops of gravitons.

The coefficient $c_\Upsilon$ is ultraviolet divergent and, using
``modified NDA'', we estimate
\beq
c_\Upsilon^{\rm NDA} = \frac{c_\delta^2}{\ell_4\ell_\delta^2}
\frac{\Lambda_{\rm NDA}^{2+2\delta}}{M_D^{4+2\delta}}
=\frac{\pi^{\delta -2}}{16 \Gamma^2 (\delta /2)}
\frac{\Lambda_{\rm NDA}^{2+2\delta}}{M_D^{4+2\delta}}.
\label{esups}
\eeq

An explicit calculation (outlined in the Appendix) gives
\beq\label{eq:c6}
c_\Upsilon =  \frac{15}{64}\frac{c_\delta^2}{\ell_4 {M}_D^{4+2\delta}} 
\int_0^\infty dk^2~k^4~G^2(k,r).
\eeq
Taking $r=0$ and introducing an explicit cut-off function $f$ we get
\beq
\frac{c_\Upsilon}{c_\Upsilon^{\rm NDA} }
\left( \frac{\Lambda_{\rm NDA}}{\Lambda}\right)^{2\delta+2}
= \frac{15}{64}
\int_0^\infty dx f(x) \int_0^\infty dy f(y) \int_0^\infty dz f(z)
\frac{z^2(xy)^{\frac{\delta}{2}-1}}{(x+z)(y+z)},~~~
x\equiv \frac{m_1^2}{\Lambda^2},~  y\equiv \frac{m_2^2}{\Lambda^2}, ~ 
z\equiv \frac{k^2}{\Lambda^2}.
\label{labit}
\eeq
The three integrations correspond to the summations over the Kaluza-Klein
modes of the two graviton propagators (with masses $m_1$ and $m_2$,
respectively) and to the 4-dimensional loop with internal momentum $k$.
\begin{itemize}
\item[a)]
Choosing $f(x)=e^{-x}$, we find that $c_\Upsilon = c_\Upsilon^{\rm NDA}$
for the values of $\Lambda/\Lambda_{\rm NDA}$ listed in the first row of table~\ref{tab4}.


\item[b)]For the case of the sharp cut-off,  $c_\Upsilon = c_\Upsilon^{\rm NDA}$
for the values of $\Lambda/\Lambda_{\rm NDA}$ listed in the second row of table~\ref{tab4}.

\item[c)] Finally, the coefficient of the single terms in $\Upsilon$
that involve fermions separated by a distance $r=1/\Lambda\ll R$ is 
(with the usual abuse of notation)
\beq
{c_\Upsilon \over c_\Upsilon^{\rm NDA} }
\left( \frac{\Lambda_{\rm NDA}}{\Lambda}\right)^{2\delta+2}
= \frac{15\delta^2(\delta +2)}{(\delta +1)(\delta +3)}2^{2\delta -7}
\Gamma^4(\delta /2).
\eeq
This gives $c_\Upsilon = c_\Upsilon^{\rm NDA}$
for the values of $\Lambda/\Lambda_{\rm NDA}$ listed in the third row of table~\ref{tab4}.

\end{itemize}
The axial-vector interaction in eq.~(\ref{upsilon}) does not affect
atomic parity violation nor $\mu$ decay and  
it is not constrained by high-precision LEP1 and SLD data.
The most stringent limits on the coefficient of the operator $\Upsilon$
come from study of contact interactions at LEP, from
dijet and Drell-Yan production at the Tevatron, from $ep$ scattering at HERA,
from neutrino-nucleon scattering.
The present bounds are summarized in 
table~\ref{tabdim6}\footnote{The terms in $\Upsilon$ and $\tau$
involving different fermions could be suppressed if
 the wave functions of fermions were significantly
splitted in the extra dimensions.
In table~\ref{tabdim6} we have separately shown the
limit from Bhabha scattering to stress that a strong bound exists also
for operators involving only electron fields, which would not be suppressed and
are not regularized
by the ``splitting'' cut-off. The data in the first two lines of
table~\ref{tabdim6} are discarded from our combination.}.
\begin{table}
\centering
\begin{tabular}{|c|c|cc|}
\hline\hline
Experiment&Process & + & $-$\\ 
\hline
LEP combined~\cite{lepcomb} & $e^+e^-$ & 11.3 & 11.5 \\
LEP combined~\cite{lepcomb} & $\mu^+\mu^-$ & 16.4 & 12.7 \\
LEP combined~\cite{lepcomb} & $\ell^+\ell^-$ & 17.2 & 15.1 \\
LEP combined~\cite{lepcomb} & $b\bar b$ & 15.3 & 11.5 \\
H1~\cite{h1} & $e^+p$ and $e^-p$ & 2.5 & 3.9\\
ZEUS~\cite{zeus} & $e^+p$ and $e^-p$ & 4.6 & 5.3\\
D\O~\cite{talk} & dijet & 3.2 & 3.1 \\
D\O~\cite{talk} & dielectron & 4.7 & 5.5 \\
CDF~\cite{talk} & dilepton & 4.5 & 5.6 \\
CCFR~\cite{ccfr} & $\nu N$ scattering & 3.7 & 5.9 \\
\hline
\multicolumn{2}{|c|}{combined}&20.6&15.7\\
\hline\hline
\end{tabular}
\caption{{\em Dimension 6 operator $\Upsilon$}:
{\em 95\% CL limits on $|c_\Upsilon/4\pi|^{-1/2}$ (in {\rm TeV}) for positive
and negative values of $c_\Upsilon$,
from different
experiments.\label{tabdim6}
}}
\end{table}

\bigskip

Graviton loops can also generate some new dimension-6 operators involving
the Higgs doublet $H$, which have the general structure
\beq
\Lag_{\rm int} = cH^\dagger H \Lag_{\rm SM},
\label{higgs}
\eeq
where $\Lag_{\rm SM}$ represents any of the SM dimension-4 interaction
terms. All coefficients $c$ are zero unless we introduce a linear
coupling between the 4-dimensional Ricci scalar and the Higgs bilinear,
as considered in ref.~\cite{radion}. On general grounds, one
expects that such coupling is non-vanishing, and
modified NDA gives the estimate $c\sim c_\Upsilon$\footnote{An explicit 
computation of the coefficients $c$ of the operator $
H^\dagger H(-\frac{1}{4}F_{\mu\nu}
F^{\mu\nu})$ discussed below, using the three regulators introduced in 
sect.~\ref{2} gives $c=0$,
because of a  cancellation between the relevant Feynman diagrams.
Such cancellations are related to the fact that the trace of the energy-momentum tensor
of massless spin-one particles vanishes in 4 dimensions.
Other regularizations can give $c\neq 0$.
We will encounter a similar ambiguous result in the next section when studying the
corrections to the anomalous magnetic moments.}.

The terms in eq.~(\ref{higgs}),
with $H$ replaced by its vacuum expectation value $v$, can be absorbed
in a redefinition of the fields and the coupling constants. Once this
is done, the interactions of the physical Higgs boson $h$ ($H\to
(v+h)/\sqrt{2}$) in a linear expansion are given by
\beq
\Lag_{\rm int} = \Lag_{\rm SM}(h) +cvh \Lag_{\rm SM}(v).
\eeq
This modifies the Higgs couplings to fermions and weak gauge bosons by a
factor $1+cv^2$. The bounds on $c_\Upsilon$ presented
in table~\ref{tabdim6} imply that this modification differs from 1 by
at most $3\times 10^{-3}$. 

More interesting is the case of operators coupling the Higgs fields
to photons and gluons 
$ {\cal O}_{h\gamma\gamma}=h \, F_{\mu\nu}F^{\mu\nu}$ and
$ {\cal O}_{hgg}=h\,G^a_{\mu\nu}G^{a\mu\nu}$,
since the graviton-mediated interaction
has to compete with a loop-induced SM term. We find that the coefficients of
these effective operators are
\beq
g_{h\gamma\gamma}=\frac{47\alpha I_\gamma}{72\pi v}-\frac{cv}{4},\qquad
g_{hgg}=-\frac{\alpha_s I_g}{12\pi v}-\frac{cv}{4},
\eeq
where $I_\gamma$ and $I_g$ are the ordinary SM loop functions~\cite{ISM}.
We have normalized them such that,
neglecting higher order corrections, $I_\gamma = I_g =1$ for $m_h \ll m_t,m_W$,
while $I_\gamma = 1.18$ and $I_g = 1.03$ for $m_h = 115\GeV$.
In fig.~\ref{fig:higgs} we show the quantity $\sigma (gg\to h)\times {\rm BR}(h\to
\gamma \gamma)$ in units of the SM value, as a function of the Higgs
mass $m_h$ and the new-physics scale $\Lambda \equiv \pm |c/(4\pi)|^{-1/2}$. Notice that
the interference of the new-physics and SM contributions in
$g_{h\gamma\gamma}$ and $g_{hgg}$ is always constructive for one coupling
and destructive for the other one, leading to a partial compensation
in $\sigma (gg\to h)\times {\rm BR}(h\to
\gamma \gamma)$. This is because the dominant SM contribution to 
$g_{h\gamma\gamma}$ is from $W$ exchange, while the one to $g_{hgg}$
is from top exchange, and they have opposite sign. Experiments at the
LHC can determine the quantity $\sigma (gg\to h)\times {\rm BR}(h\to
\gamma \gamma)$ with a precision of 10--15{\%}~\cite{zepp}, and therefore
significantly probe the virtual-graviton contribution, as apparent from 
fig.~\ref{fig:higgs}.
The implications
of the new couplings $g_{h\gamma\gamma}$ and $g_{hgg}$ have also
been considered in ref.~\cite{hall}.



\section{Gravitons and SM gauge bosons}

Although the operators $\tau$ and $\Upsilon$ do not directly affect
precision electroweak data, they contribute when they are dressed by
gauge bosons. This corresponds to
loops with exchange of virtual gravitons and vector bosons 
(such those illustrated in fig.~\ref{fig:treeloop}c) leading to
new dimension-6 operators~\cite{Contino}.
We estimate them by making the conservative assumption that the SM 
gauge couplings remain weak up to 
the cutoff.

Since graviton loops are flavour universal (neglecting the bottom 
quark mass)
gravitational corrections to the various electroweak precision
measurements can be embedded in three parameters
that are usually chosen to be 
$\epsilon_1,\epsilon_2,\epsilon_3$~\cite{epsilon}\footnote{Graviton 
corrections to the $S,T,U$ parameters~\cite{STU} are plagued by 
large gauge-dependent infrared effects
that, in the unitary gauge, unphysically increase with increasing $M_D$.
The reason is that $S,T,U$ parameterize new physics present only in the
vector boson sector and therefore are not physical observables in the 
case of gravity, which couples to everything.}. Similarly, 
we can also estimate the contribution to the anomalous magnetic 
moment of the muon.
The results from ``modified NDA'' are
\begin{eqnarray}
\delta\epsilon_i^{\rm NDA} &\approx & \frac{c_\delta M_Z^2\Lambda^\delta_{\rm NDA}}{\ell_4 \ell_\delta
M_D^{\delta +2}}=\frac{\pi^\frac{\delta -4}{2}}{16\Gamma(\delta/2)}
~\frac{M_Z^2\Lambda^\delta_{\rm NDA}}{M_D^{\delta +2}}, \\
\delta a_\mu^{\rm NDA} &\approx & \frac{c_\delta m_\mu^2
\Lambda^\delta_{\rm NDA}}{\ell_4 \ell_\delta
M_D^{\delta +2}}=\frac{\pi^\frac{\delta -4}{2}}{16\Gamma(\delta/2)}
~\frac{m_\mu^2\Lambda^\delta_{\rm NDA}}{M_D^{\delta +2}}.
\label{amu}
\end{eqnarray}
Because of a cancellation between different Feynman diagrams, 
the one-loop graviton correction to $\delta a_\mu$ is zero when
estimated with the regulators discussed in sect.~\ref{2}~\cite{Contino}.
A finite result in agreement with the estimate in eq.~(\ref{amu}) is obtained
using other regularizations, like the non-supersymmetric dimensional 
regularization~\cite{grasser}
that spoils the cancellation by acting in
a different way on fermions and vector bosons.


In our analysis, we will take as a representative bound from electroweak
data $|\delta \epsilon_i|<10^{-3}$. Then, eq.~(\ref{amu}) implies
$|\delta a_\mu |<10^{-9}$, which is about 1 $\sigma$ of the experimental
value and of the SM theoretical estimate. Therefore, electroweak data
and the anomalous magnetic moment of the muon give comparable bounds on
virtual graviton effects.

\section{Comparison of the different constraints}

In this section we attempt a comparison between the constraints
from the different collider observables. 
For virtual-graviton effects, we will
use the estimates from ``modified NDA'', and then comment on other
regulators. 
Once again, we stress that this procedure does not predict
order-one coefficients and that, in this sense, our bounds are 
only semi-quantitative.

Notice also that the dominant effects at colliders are produced 
by the heaviest Kaluza-Klein modes allowed by the kinematics of the relevant
process. On the other hand,
the lightest Kaluza-Klein modes 
affect astrophysical processes, giving bounds that for $\delta = 2,3$
are much stronger than collider bounds~\cite{SN1987+graviton}.
In particular, supernova bounds directly exclude observable
deviations from the Newton law at sub-millimeter scales
caused by extra-dimensional gravitons.
Reasonable modifications of the low-energy part of the spectrum
may remove the lightest Kaluza-Klein modes, without affecting
the high-energy signals that we consider.

\smallskip

Our results are summarized in fig.~\ref{fig:vincoli}. 
The scale $\Lambda$ that we have introduced in this analysis 
plays an important r\^ole in comparing the different observables,
since it parametrizes the relative strength of tree-level versus
loop effects or, in other words, it defines how strongly-interacting 
gravity is. 
Setting $\Lambda = M_D$, as  assumed in many analyses,
is an arbitrary restriction: $\Lambda$ is a relevant free parameter.

As apparent from fig.~\ref{fig:vincoli},
the most stringent bounds arise from graviton emission and from 
the dimension-6 operator $\Upsilon$, which is severely constrained
by the recent LEP2 data.
The limit on the operator $\tau$ is less significant than the bound from
$\Upsilon$ at large values of
$M_D$, and weaker than the bound from graviton emission at small $M_D$.
However, it rules out a small additional region of parameter space
at intermediate values of $M_D$.
The LEP2 bounds on $\Upsilon$ give a stronger constraint  
than electroweak precision measurements or $(g-2)_\mu$.

\smallskip

In order to study the significance of the ${\cal O}(1)$
factors not controlled by NDA, in the previous sections we have
computed the coefficients of the relevant operators with
three different arbitrary cut-off procedures.
For $\Lambda =\Lambda_{\rm NDA}$,
regulators with an exponential functional behavior or
a sharp cut-off give smaller values of $c_\Upsilon$ than the ``modified
NDA'' estimate, while the splitting regulator gives larger values.
However, these factors can be reabsorbed in the definition of $\Lambda$.
For each regularization procedure,
we choose to define the ratio $\Lambda/\Lambda_{\rm NDA}$ 
by enforcing $c_\Upsilon = c_\Upsilon^{\rm NDA}$,
separately for all values of $\delta$. The results are shown in 
table~\ref{tab4}. 
With this definition, the bounds from the $\Upsilon$ operator,
using explicit regulators,
are equal to those plotted in fig.\fig{vincoli}.
The bound from graviton emission (vertical lines in fig.\fig{vincoli})
does not depend on $\Lambda$, and therefore remains unmodified.
The coefficient of the $\tau$ operator is then determined
and its value in units of $c_\tau^{\rm NDA}$ is given in table~\ref{tab5},
for the different regulators. 
Since $c_\tau$ scales as $M_D^{-4}$,
for all the considered cut-off procedures
the bounds from $\tau$ are only slightly 
stronger than in the NDA case, shown in fig.\fig{vincoli}.
In this respect, once the different cut-off parameters
are appropriately compared, our results are rather insensitive
of the regularization procedure. 
We do not need to discuss how the other bounds change, since they are always
sub-dominant.

\begin{table}
\centering
\begin{tabular}{|cc|ccccc|}
\hline\hline
&$\Lambda/\Lambda_{\rm NDA}$ &  $\delta = 2$&  $\delta = 3$&  $\delta = 4$&
  $\delta = 5$&  $\delta = 6$ \cr
\hline
\hbox{a)}&\hbox{exponential cut-off} &1.57  &1.51&1.40&1.29&1.19\cr
\hbox{b)}&\hbox{sharp cut-off} &1.58&1.60&1.56&1.52&1.48\cr
\hbox{c)}&\hbox{splitting cut-off} &0.89&0.76&0.64&0.55&0.47\cr
\hline\hline
\end{tabular}
\caption{\label{tab4}
{\em The values of $\Lambda/\Lambda_{\rm NDA}$ in different
regularization procedures obtained by imposing $c_\Upsilon =
c_\Upsilon^{\rm NDA}$.
}}
\end{table}

\begin{table}
\centering
\begin{tabular}{|cc|cccc|}
\hline\hline
&$c_\tau/c_\tau^{\rm NDA}$ &    $\delta = 3$&  $\delta = 4$&
  $\delta = 5$&  $\delta = 6$ \cr
\hline
\hbox{a)}&\hbox{exponential cut-off}  &2.68&1.97&1.91&2.04\cr
\hbox{b)}&\hbox{sharp cut-off} &3.20&2.45&2.34&2.37\cr
\hbox{c)}&\hbox{splitting cut-off} &2.40&1.66&1.55&1.61\cr
\hline\hline
\end{tabular}
\caption{\label{tab5}
{\em The values of $c_\tau/c_\tau^{\rm NDA}$ in different
regularization procedures obtained by imposing $c_\Upsilon =
c_\Upsilon^{\rm NDA}$. For $\delta =2$ the comparison is affected by 
the infrared divergence, since $c_\tau /c_\tau^{\rm NDA}=\ln (\Lambda^2/E^2)$.
However, this is also of order unity for relevant values of the typical
energy $E$.
}}
\end{table}


\medskip

The limits on $\Upsilon$ from LEP2 imply that the gravitational
theory can never become strongly interacting, in the experimentally 
relevant range of $M_D$, since the solid line in fig.\fig{vincoli} 
corresponding
to $\Lambda =\Lambda_{\rm S}$ lies in the excluded region. 
A low cut-off $\Lambda$, necessary to reduce 
the contributions from graviton loops, has important implications for
collider experiments. It reduces the sensitivity region from graviton 
emission, but it opens up the possibility of discovering the new states
which presumably lie at the scale $\Lambda$ and are responsible for
the softening of the ultraviolet behaviour of gravitational interactions.
Especially if $\Lambda \ll M_D$, the new physical phenomena related to
the scale $\Lambda$ could give the discovering signatures
in present and future experiments.
Ignoring them is clearly a strong limitation of graviton phenomenology.

The physics at the scale $\Lambda$ could itself generate new effective
operators, potentially giving stronger constraints than those 
considered here.
One possibility is that $\Lambda$ be the scale of some string model,
and the interaction strength 
be related to the string coupling.
Although one cannot make model-independent statements, it is plausible
to expect that $\Lambda$ has to be larger than few TeV.
Alternatively, $\Lambda$ could be the mass scale of supersymmetric particles.
Assuming conserved matter parity, physics at the scale $\Lambda$ generates
effective dimension-6 operators
only at one-loop order, and the lower bounds on $\Lambda$ are very weak,
of the order of $100 \GeV$.

\medskip

Finally, we recall that a reparametrization invariant extra-dimensional
theory containing SM fields localized on a 3-dimensional brane
contains not only gravitons but also
brane fluctuations, described by $\delta$ neutral scalars named `branons'.
This can be understood geometrically: a brane straight 
in some system of coordinate is bent when described using different
coordinates.
(An exception is when the brane is located at special singular 
points of the extra dimensional space;
such pathological spaces are often studied because of their mathematical 
simplicity).
Branon couplings are controlled by the tension $\tau$ of the brane,
just like graviton couplings are controlled by the Planck mass $M_D$.
Therefore, just like it is possible to study effects generated by gravitons alone,
one can also study effects generated by branons alone.
Colliders provide the most stringent constraints. 
With $\delta$ flat extra dimensions
branon missing energy signals are equal to missing energy signals produced by gravitons in 6 extra dimensions,
rescaled by the factor $\pi \delta c_{6} M_{10}^8/30\tau^2$
(predicted to be 1/5 by the simplest toy string models)~\cite{branons6}.
The same rescaling factor applies to branon virtual effects:
tree level (one loop) graviton exchange corresponds to one loop (three loops) branon exchange.
Therefore fig.\fig{vincoli}d
 also describes branon phenomenology
after reinterpreting $M_D\to\tau$ as described above.
If $\tau\ll M_D^4$ brane fluctuations suppress graviton effects~\cite{recoil}.

\section{Conclusions}

Collider experiments probe theories with extra-dimensional gravity in
a model-independent way through graviton emission, and in a model-dependent
way through studies of contact interactions. We have given some tools that
allow for a comparison between the two, under the assumption that the
coefficients of the new contact interactions are dominated by graviton
effects (at tree or loop level). It is important to emphasize that our
analysis has two important limitations: {\it i)} the coefficients of
the effective operators induced by graviton exchange (at tree or loop level)
can only be estimated, since they are sensitive to the ultraviolet;
{\it ii)} because of this sensitivity, other contributions from unknown
new physics can be equally or more important.

Nevertheless, we believe that some interesting results have been reached
by this analysis. Because of the different dimensionality of the operators
involved, graviton loop corrections can be more effective than tree-level 
exchange in constraining the theory from present data. We have found
that the structure of effective operators generated by graviton loops
is relatively simple. At the level of dimension-6 operators, loop diagrams
with an intermediate state of virtual gravitons generate only the
operator $\Upsilon$ and some operators relevant for Higgs physics.
Mixed loops with graviton and gauge bosons intermediate states can
generate other operators which affect electroweak precision measurements,
but which are less effective in constraining the theory. Therefore global
analyses
of extra-dimensional gravitons at colliders should simultaneously take
into account graviton emission and the effects of the operators $\tau$
and $\Upsilon$.

LEP2 data constrain the coefficient of the operator $\Upsilon$ and thus
put one of the strongest limit on the existence of extra-dimensional
gravity at the weak scale. In particular, they disfavour strongly-interacting
gravity at accessible energies. This has important implications for
future searches at high-energy colliders, limiting the available parameter
space where graviton emission can be observed, but leaving open the
possibility of detecting the physics responsible for the premature
softening of quantum gravity.

\paragraph{Acknowledgments}
We wish to thank R. Contino, L. Pilo and R.~Rattazzi for useful discussions.

\appendix

\section{Graviton-fermion loops}
In  order to write reparametrization invariant Lagrangians involving fermions
we introduce the $D$-bein $E^A_{M}$ relating generic coordinates $x_M$ to
locally  free-fall systems $\xi_A$, so that
the metric is given by $g_{MN} = \eta_{AB} \, E^{A}_{M}  E^{B}_{N}$.
The $D$-bein basis definition introduces an additional gauge symmetry,
besides diffeomorphisms,  due  to the freedom of rotating  $\xi_A$ with a local Lorentz transformation. 
With a clever gauge choice~\cite{des}
Lorentz ghosts are absent  and the $D$-bein can be reexpressed in terms of the metrics.
Expanding the metrics in fluctuations around flat space, $g_{MN} = \eta_{MN} + \kappa h_{MN}$
one has ($\kappa^2 = 4 c_\delta /M_D^{2+\delta}$)
\beq\label{eq:eAM}E^A_M = \delta_{AM} + \frac{ \kappa }{2}h_{AM} -\frac{\kappa^2}{8} (h\cdot h)_{AM}+
\frac{\kappa^3}{16} (h\cdot h\cdot h)_{AM}+\cdots\eeq
This  gauge choice can be generalized to systems involving branes~\cite{Contino}
in such a way that the brane vierbein $e^a_m$ keeps the same form as 
$E^A_M$ in eq.\eq{eAM}
with the metrics replaced by the induced metrics on the brane:
\begin{equation}
h_{MN}(x^R)\to h_{mn}(x^r) = h_{mn}^0 +\frac{1}{\kappa}
(\partial_m \xi_i) (\partial_n \xi^i)+
(\xi^i \partial_i h_{mn}^0+ h_{im}^0 \partial_{n} \xi^i+ h_{in}^0 \partial_{m} \xi^i)+\cdots
\end{equation}
where $h^0(x^r)$ denotes $h(x^r, x^i = 0)$, the $D$-dimensional graviton field evaluated at the brane rest position $x^i=0$.
We have splitted $D$ dimensional indices $M,N$ into their four-dimensional components $m,n$
plus their extra dimensional components $i,j=\{1,\ldots,\delta\}$.
The `branons' $\xi^i(x_m)$ are $\delta$ scalar fields  that fix the position of each point of the brane
in the extra dimensions as $x^i(x_m) = \xi^i(x_m)$.
This choice completely fixes brane reparametrizations without giving  ghost fields~\cite{sundrum}. 
The kinetic term of a Dirac field localized on a brane with tension $\tau$ is 
\beq 
S = \int d^4 x~e~\bigg[-\tau + e^m_a \frac{i}{2}[\bar{\Psi}    \gamma_a  (D_m \Psi) - (\overline{D_m  \Psi})  \gamma_a  \Psi]\bigg],
\qquad
D_m = \partial_m + \frac{1}{8} [\gamma_a,\gamma_b]\omega_m{}^{ab},
\eeq
where $e=\det e_m^a$ and the graviton contribution to the spin-connection is
$$
\omega_m{}^{ab} =
\frac{\kappa}{2} \partial_b h_{am}+\frac{\kappa^2}{8}(h_{bn}\partial_m h_{an}+
2h_{an}\partial_n h_{bm} + 2h_{bn}\partial_a h_{mn}) - (a\leftrightarrow b) +
\cdots
$$
By expanding the action up to second order in $\kappa$ one finds the off-shell
$\bar\Psi \Psi h$ and $\bar\Psi \Psi hh$ couplings.
Taking into account that Feynman graphs with branons or with 
corrections to fermionic gravitational vertices give no one-loop 
contribution to the dimension-6 operator $\Upsilon$,
we only need to compute the following graphs:
$$\includegraphics[width=17cm]{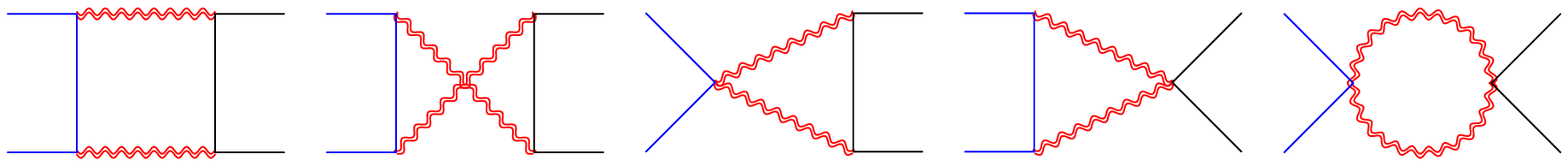}$$
We computed them using two different gauges for the graviton KK,
the deDonder and the unitary gauge (see ref.~\cite{Contino}), 
verifying that their sum
is gauge invariant mode by mode
(so that we can perform this check without introducing any regulator).
The result is given by eq.\eq{c6}.
It contains no factor $\delta$ because the $\delta$-dependent `scalar'
part in the graviton/radion propagator gives no contribution.


%

\frenchspacing
\begin{multicols}{2}

\end{multicols}
\end{document}